# An Improved WBF Algorithm for Higher-Speed Decoding of LDPC Codes

MA Ke-xiang, LI Yong-zhao, ZHU Cai-zhi, ZHANG Hai-lin and QI Feng

*(The National Key Laboratory on ISN, Xidian University, Xi'an 710071, China)*

**Abstract:** Due to the speed limitation of the conventional bit-chosen strategy in the existing weighted bit flipping algorithms, a high-speed LDPC decoder cannot be realized. To solve this problem, we propose a fast weighted bit flipping (FWBF) algorithm. Specifically, based on the stochastic error bitmap of the received vector, a partially parallel bit-choose strategy is adopted to lower the delay of choosing the bit flipped. Because of its partially parallel structure, the novel strategy can be well incorporated into the LDPC decoder [1]. The analysis of the decoding delay demonstrates that, the decoding speed can be greatly improved by adopting the proposed FWBF algorithm. Further, simulation results verify the validity of the proposed algorithm.

**Key words:** stochastic error bitmap; bit-choose strategy; partially parallel; pipelined.

## I. Introduction

Low-density parity-check (LDPC) codes [2], were first proposed by Gallager in the 1960s and later resurrected by MacKay and Neal [3]. LDPC has attracted considerable attention due to its near Shannon-limit performance and inherently parallelizable decoding structure [4] [5]. There have been two main kinds of algorithms to decode the LDPC codes, BP-based algorithm and BF-based algorithm. BP-based algorithms have been extensively applied for its great decoding performance, especially normalized min-sum algorithm and offset min-sum algorithm. However, because of the huge check-parity matrix of the LDPC codes, a BP-based decoder results in huge hardware expenditure. Furthermore, due to the limitation of the hardware resource, especially the memory resource [6], almost LDPC decoders adopt a partial parallel structure, which limits the decoding speed to a narrow range. Consequently, the BP-based algorithms are unsuitable for fast decoders. However, with the increase of data services, the communication systems require much higher data transmission speed. For example, for the next generation of optical communications systems operating at 40 Gb/s, the current circuit technology does not accommodate BP-based decoding and only BF-based decoding is possible [7].

Because of the shortcomings of BP-based algorithm mentioned above, BF-based algorithm has abstracted extensive study in recent year. Kou provided weighted BF (WBF) [5], which first bridged the performance gap between BP-based algorithm and BF-based algorithm, and a series of improved algorithms are proposed to further improve the decoding performance of WBF algorithm, defined as WBF-based algorithms. By incorporating the received value into the metric value, J. Zhang presented a modified WBF (MWBF) algorithm [8]. In terms of the update of the check message in the min-sum algorithm, an improved MWBF (IMWBF) algorithm was further proposed [9]. To improve the convergence speed of IMWBF algorithm, X. Wu proposed a parallel version of the IMWBF algorithm (PWBF) [10]. Further, an approach is proposed for PWBF algorithm to obtain the optimal thresholds [11]. On the other hand, introducing a fully different metric value, Liu and Pados proposed a new WBF algorithm, defined as LP-WBF [12]. Like the PWBF, multi-bit LP algorithm is proposed to improve the decoding speed [13], defined as MLP-WBF in this paper.

This work was supported in part by the National Science Foundation of China (No.61072069), the 111 Project (B08038).

As is discussed above, because of their low computation complexity, BF-based algorithms can be utilized to design a high-speed LDPC decoder, and two multi-bit flipping algorithms are adopted to improve the convergence speed obviously [11] [13]. In the multi-bit flipping algorithms, it is critical to choose the bits flipped in every iteration. The method proposed by Wu is efficient to obtain well decoding performance. However, its complex structure limits the decoding speed of the LDPC decoder. MLP-WBF has a simple method to choose the flipped bits. However, the process of the bits choose can only been executed sequentially, which also limits the decoding speed. For further improving the decoding speed of the LDPC decoder, we propose a partially parallel choose mechanism, which utilizes the received vector with stochastic distribution noise. Correspondingly, a new WBF-based algorithm with low decoding delay is proposed, defined as fast WBF (FWBF) algorithm.

The remainder of the paper is organized as follows. In Section II, we introduce the basic notation and provide the required background material. In Section III, a fast weighted bit flipping is proposed, which has a quite low decoding delay. Section IV provides a detailed analysis of the decoding delay of the proposed algorithms. We illustrate numerical results in Section V, and this paper is concluded in the last section.

## II. Conventions and Preliminaries

An LDPC code is a linear block code given by the null space of a "sparse" parity-check matrix $H$, and "sparse" means that non-zero entries are far more than the zero entries in the parity-check matrix. A LDPC code is defined as a $(r,q)$-regular code if its parity-check matrix has exactly $r$ "non-zero" entries in each row and $q$ "non-zero" entries in each column. For almost of LDPC codes, the following constraint is imposed on the rows and columns of the parity-check matrix $H$ of a code: no two rows (or two columns) can have more than one place where they both have "non-zero" entries, i.e., the famous row-column (RC) constraint. Let C be a $(N,K)$ LDPC code of block length $N$ and dimension $K$, which has a parity-check $H = [h_{MN}]$ of $M$ rows, and $N$ columns. Let $R_c = K/N$ denote its code rate. In this paper, we assume BPSK modulation, which maps a code $c = [c_1 \ c_2 \ \cdots \ c_N]$ into a transmitted sequence $s = [s_1 \ s_2 \ \cdots \ s_N]$, according to $s_n = 1 - 2c_n$, for every $n \in (1,2\ldots N)$. Then, $s$ is transmitted over an AWGN channel, and $y = [y_1 \ y_2 \ \ldots \ y_N]$ is the received codeword corresponding to the transmitted sequence $s$. $z_k \ (0 < k \leq K_{max})$ represents the tentative binary hard-decision vector at the end of the k-th decoding iteration and $K_{max}$ is the maximum number of the decoding iteration. The weight of the syndrome vector $s_k = z_k \times H^T$ is denoted by $w_k$. Let $\mathcal{M}(n) = \{m : h_{mn} = 1\}$ denote the set of check nodes including the bit node $n$, namely, the position of ones in the $n$-th column of $H$; the set of bit nodes participating in check node $m$ is denoted by $\mathcal{N}(m) = \{n : h_{mn} = 1\}$, namely, the position of ones in the $m$-th row of $H$. In addition, $\mathcal{M}(n) \backslash m$ represents the set $\mathcal{M}(n)$, excluding the $m$-th check node and $\mathcal{N}(m) \backslash n$ represents the set $\mathcal{N}(m)$, without the $n$-th bit node.

In IMWBF algorithm [9], the metric value of the *n*-th bit node is computed as follows,

$$E_n = \sum_{m \in \mathcal{M}(n)} (2s_m - 1)w_{n,m} - \alpha |y_n| \quad (1)$$

where $w_{n,m} = \min_{i \in \mathcal{N}(m) \setminus n} |y_i|$. At each iteration, the bit node with the largest metric value $E_n^{MAX}$ will be flipped.

### III Proposed Algorithm

In this section, based on IMWBF algorithm, we propose a fast weighted bit-flipping algorithm. In order to show the advantages of such algorithm by corresponding analysis, the existing PWBF algorithm and MLP-WBF algorithm are introduced as a comparison. PWBF algorithm is a multi-bit flipping algorithm of IMWBF algorithm, which has a well decoding performance and convergence speed. However, because of its complex calculation process, the bit-choose strategy in PWBF algorithm is hard to be implemented with hardware under certain condition. Therefore, the MLP-WBF algorithm with a simple bit-choose method in is proposed as a proper alternative of PWBF algorithm in this manuscript, which is more convenient to design a LDPC decoder and has a satisfied decoding performance and convergence speed as well.

In the MLP-WBF, the steps of choosing the bits with the $\lambda$ largest metric value to flip can be merely executed sequentially. Furthermore, due to the limitation of the hardware resource, the choosing for each flipped bit can only be realized in partially parallel. Consequently, the choosing of the flipped bits in MLP-WBF will lead to a considerable decoding delay, and cannot satisfy the demand of the high-speed communication. The drawback of MLP-WBF mentioned above also represents the main disadvantages of existing PWBF algorithm. The FWBF proposed in this paper will solve the problems effectively, which is discussed by contrasting to MLP-WBF in detail in following paragraphs.

When the signal is transmitted in the AWGN channel, error bits will be randomly uniformly distributed in the received vector. If it is transmitted in the fading channel, burst error will appear. Consequently, burst error will result in too much contiguous error bits in one received vector, which may exceed the decoding ability of error-correct codes. However, almost all the communication systems transmitted in the fading channel adopt interleavers to randomize noise, for example IEEE 802.16 [14]. Correspondingly, the error bits will be uniformly distributed in every received vector, and then the received vectors with burst error can be correctly decoded. Therefore, we can suppose that all the received vectors sending to the decoder have a uniformly distributed error bits in most communication systems. Therefore, when the codeword is divided into several blocks, it is reasonable to consider that error bits are uniformly distributed in every block. Combined with the common structure of the LDPC decoder [15], we propose a partially parallel bit choose algorithm. Specifically, in the process of the iterative decoding, we divide the metric value vector into several blocks, and then choose one bit to flip in every block at most. The decoding performance of FWBF is similar to that of MLP-WBF, and its significant advantage in decoding speed makes it a better choose in the high-speed communication systems. The proposed algorithm (FWBF) is described as follow:

*Initialization*: Set the iterative number $k = 0$, and let the initial decision vector $z_0 = (1 - \text{sgn}(y))/2$; For each check node *m*, $0 \leq m < M$, compute $w_{n,m}$ for every bit nodes in it.

*Step*1: compute the weight $w_k$ of syndrome vector. If ($w_k = 0$ or $k > K_{max}$), stop the decoding process and output $z_k$;

*Step*2: for each *p*-length block $i$ ($i \in [1, N/p]$), compute the metric values of all the bit nodes in the *i*-th block by Eq. (1), and then find the bit node *b* which has the biggest metric value in the current block; if ($E_n^b > 0$) flip $z_b$.

$k= k+1$, go to *Step* 1.

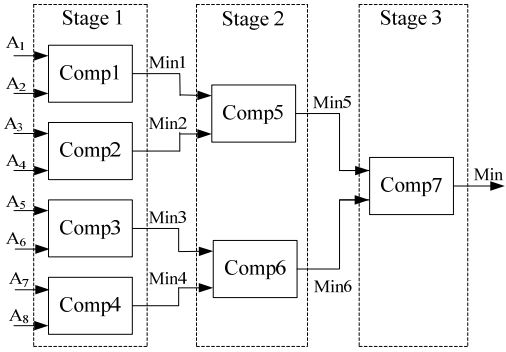

Fig. 1. 3-*stage* comparator network

### IV. Delay Analysis

The LDPC decoders are commonly implemented by the partially parallel structure. In this paper, we also utilize this structure to discuss delay performance of the FWBF and MLP-WBF. Suppose that both two algorithms have the same basic decoding unit (calculate $E_n$), and $2q$ metric values can be calculated in parallel. Correspondingly, they both require $N/(2q)$ clk to calculate all the metric values. Besides, we also utilize the comparator network with the similar structure (Fig. 1) to implement the bit choose for both the algorithms. In this comparator network, there are $q = 4$ comparators at the first stage, which can simultaneously implement the comparison between $2q = 8$ metric values, and its comparators are divided into $\alpha = log_2(2q) = 3$ stage. Correspondingly, after inputted $2q = 8$ metric values, comparator network outputs their largest value after $log_2(2q) = 3$ clk. Because of the pipelined structure of the comparator network, the intermediate data of different block simultaneously goes down the $\alpha$-stage comparator network by clock. For example, the data of the first block is operated in the first stage in the first clk, and its obtained results are inputted into the second stage in the second clk. Meanwhile, the data of the second block enters in the first stage. Therefore, for the general case, after $log_2(2q) + N/(2q)$ clk, all the local maximum values of every block can be figured out, and then additional $log_2(N/2q)$ clk is required to obtain the maximum of the $N/(2q)$ local maximum values, i.e., the global maximum. In MLP-WBF, the different maximum can only be calculated one by one. Therefore, the total decoding delay (one iteration) to calculate the $\lambda$ largest metric value is described by:

$$\lambda \times (log_2(2q) + N/(2q) + log_2(N/2q)) \quad (2)$$

For the proposed algorithm, we also utilize the same comparator network to calculate the maximum. Once obtained metric values of some block, they will be immediately inputted into the comparator network. Contributing to the pipelined structure, we only spend $log_2(2q)$ clk to obtain the maximum of the last block after its metric values were calculated. Therefore, the total decoding delay (one iteration) to choose the bits flipped in the proposed algorithm (FWBF) is:

$$log_2(2q) + N/(2q) \quad (3)$$

Taking the EG (255, 175) LDPC codes for instance, we discuss the decoding delay of the two proposed algorithms in detail. The 255 metric values are calculated by 16 parallel basic decoding units[1], and only spend 16 clk. For FWBF and MLP-WBF, there are 8 comparators in the first stage of the comparator network, separately. By Eq. 2, we calculate that the decoding delay caused by the choosing of the bits flipped is $24\lambda$ in MLP-WBF. However, for FWBF, the corresponding decoding delay is only 20. In the decoding process of MLP-WBF, the typical value of $\lambda$ is about 10. Furthermore, from the next section, it is noticed that the two algorithms have similar average iterative numbers. Therefore, we can conclude that FWBF has a significant advantage in the decoding delay, compared with MLP-WBF. Furthermore, MLP-WBF requires storing the metric value of every bit, consuming considerable memory resource. Because of the scarcity of the memory resource [16], it increases the implement hardness of MLP-WBF to a certain extent, especially for long codes.

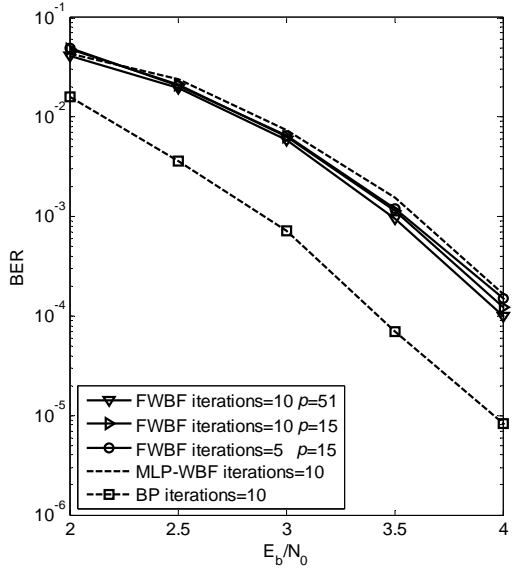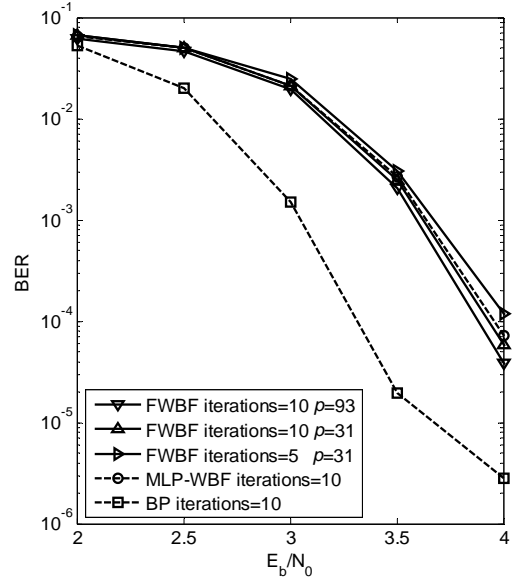

Fig. 2. Performance comparison of BP, PWBF and IPWBF decoding for EG (255,175) code

Fig. 3. Performance comparison of BP, PWBF and multi LP-WBF decoding for EG(1023,781)code

## V. Simulation Analyses

In the following, numerical results for the proposed algorithms are provided. EG (255, 175) code and EG (1023,781) are adopted [5]. For the convenience of description, we refer to them as code 1 and code 2, separately. Moreover, an AWGN channel with zero mean and variance $N_0/2$ is assumed. BPSK modulation is adopted. For each SNR, 100 codeword errors are watched.

In Fig. 2 and Fig. 3, a series of performance comparisons between the proposed algorithm and the existing algorithms are provided. We noticed that the performance of FWBF is similar to that of MLP-WBF for the two codes under different block length. For both code 1 and code 2, the decoding performance under two different kinds of block length is also provided. With the increase of the block length, decoding performance is gradually increased, but performance gap is not significant. Furthermore, the performance gap between 5 iterations and 10 iterations is not distinct for the two codes. This means 5 iterations is enough to make the proposed algorithm converge well, which is abstractive to implement a high- speed decoder. Specially, BP is 0.5 dB better than FWBF when the BER is $10^{-4}$ for code 1. However, for some communication systems, such as the satellite communications system and optical communications system, they have a well channel environment and a super high data rate. Therefore, a faster algorithm is more required than the algorithm which has a little better decoding performance. Furthermore, because of its low computational complexity,

---
[1] We introduce a "null" metric value, and then all the metric value can be divided into 16 equal blocks.

FWBF is a more comparative algorithm under those systems.

Tab. 1. The averaged iterations of MLP-WBF and FWBF decoding ($K_{MAX}$ =10 and $N$=1023).

|  | 3.0 dB | 3.5 dB | 4.0 dB | 4.5 dB |
|---|---|---|---|---|
| MLP-WBF | 8.86 | 6.20 | 4.54 | 3.83 |
| FWBF ($p$=93) | 9.29 | 7.37 | 6.06 | 5.12 |
| FWBF ($p$=31) | 8.62 | 5.60 | 4.19 | 3.64 |

In Tab 1, the average iterative numbers are provided for MLP-WBF and FWBF algorithm. With the increase of the block length, for FWBF algorithm, the averaged iterative number is gradually increased. Therefore, the secondary largest block number can obtain a faster decoder. In view of the limitation of the hardware resource, for the two codes, we also choose the secondary largest block number to design a decoder rather than the largest one, although the latter one has a little better decoding performance. Furthermore, we also notice that MLP-WBF and FWBF have similarly averaged iterative number. Provided that FWBF requires far fewer clocks to implement one iteration, a faster decoder can be constructed by this algorithm.

## VI. Conclusion

In this paper, we have proposed a fast weighted bit flipping algorithm for LDPC codes. Compared with the existing algorithm, the proposed one has similar decoding performance but very higher decoding delay. Therefore, the proposed algorithm is a better choose for the high-data communications systems. Furthermore, its inherently partially parallel structure makes it easier to implement. The validity of the proposed algorithm for different LDPC codes is verified by the simulation results.